\documentclass[%
reprint,
 amsmath,amssymb,
 aps,
 pra,
floatfix,
]{revtex4-2}

\usepackage{graphicx}
\usepackage{dcolumn}
\usepackage{bm}

\usepackage{xcolor}
\usepackage{url}
\usepackage{multirow}
\usepackage{comment}

\newcommand{\affa}{Future Research Lab, China Mobile Research Institute, Beijing 100053, China}
\newcommand{\affb}{QICI Quantum Information and Computation Initiative, Department of Computer Science, The University of Hong Kong, Pokfulam Road, Hong Kong}
\newcommand{\affc}{Department of Physics, Renmin University of China, Beijing 100872, China}

\newtheorem{theorem}{Theorem}

\newtheorem{definition}{Definition}

\usepackage{physics}

\begin{document}

\preprint{APS/123-QED}

\title{Stream randomness extraction against quantum side information}

\author{Chun-Yang Luan}
\affiliation{\affa}

\author{Cheng-Kang Pan}
\affiliation{\affa}


\author{Xiang Zhang}
\email{Corresponding author: siang.zhang@ruc.edu.cn}
\affiliation{\affc}

\author{Xingjian Zhang}
\email{Corresponding author: thuxjzhang@gmail.com}
\affiliation{\affb}

\date{\today}

\begin{abstract}

Randomness extraction is indispensable for quantum random number generators, serving to eliminate bias and potential information leakage from raw measurement data. Conventional extractors operate in a block-wise fashion, requiring the complete accumulation of raw data before processing. To circumvent the latency and buffering overheads that hinder real-time random number generation, recent work~\citep{huang2022stream} introduced a stream-cipher implementation for the randomness extractor based on the Toeplitz matrix hashing. In this work, we generalize this stream-processing paradigm to the broader family of randomness extractors based on (almost dual) universal$_2$ random hashing. Specifically, we shift the computational burden from a time-consuming block-wise post-processing stage into an offline preprocessing stage that generates a pseudo-random mask. This allows the raw data to be processed by the mask on the fly using a simple bitwise exclusive-OR operation. Crucially, we prove that this stream implementation strictly preserves the security guarantees of the original block-wise protocols. We detail the transformation of three typical constructions---based on standard Toeplitz, circulant, and modified Toeplitz matrices---from block to stream implementations, and benchmark their practical performance using realistic quantum experimental data. We anticipate our framework will enhance the efficiency of real-time quantum cryptographic systems.

\end{abstract}

\maketitle

\section{Introduction}
Randomness is a ubiquitous resource in information processing, serving as an essential ingredient for virtually all modern cryptographic primitives and a variety of computational algorithms. Unlike classical systems, quantum measurements yield intrinsically unpredictable outcomes, providing a fundamental mechanism for generating true randomness. Rapid advances in quantum information theory and experimental techniques have led to the realization of quantum random number generators (QRNGs) across a wide range of physical platforms~\cite{herrero2017quantum,ma2016quantum,mannalatha2023comprehensive}.
With a proper characterization of the underlying physical processes, the amount of intrinsic randomness can be rigorously quantified using information-theoretic entropic measures. 

Despite containing true randomness, the raw output generated by the quantum measurements cannot typically be utilized directly for information processing. Raw measurement distributions are frequently biased and, critically, may be correlated with side information accessible to an eavesdropper~\cite{ma2013postprocessing,ma2016quantum}. That is, the outcomes are not fully random and could compromise the reliability and even security of their uses. To eliminate these inherent imperfections and extract an unbiased random bitstring with (nearly) full entropy, a post-processing procedure is mandatory for the QRNG. For this purpose, leveraging the established bounds on the source's entropy, a seeded randomness extractor can be applied to the raw data \cite{carter1977universal,nisan1996randomness,bennett1988privacy,bennett1995generalized,renner2005universally,tomamichel2011leftover}. The extractor uses a short sequence of perfectly random bits---called a \emph{seed}---to power extraction functions, systematically distilling the raw, correlated outcomes into a rigorously secure resource of random numbers.

Conventionally, randomness extractors operate on blocks of data, requiring the complete generation of raw measurement outcomes before offline post-processing. 
This block-wise processing procedure may not be ideal for real-time applications.
In systems such as satellite-based quantum communication~\cite{huang2022stream,li2024improved}, data acquisition from quantum measurements is typically slow. This creates latency during which the extractor module remains idle. In addition, in protocols such as device-independent generation~\cite{pironio2010random}, the raw data typically contains low entropy. Randomness extractors are less efficient and the offline data processing becomes time-consuming. 
Stream randomness extraction was introduced to address the above issues~\cite{huang2022stream}. In randomness extractors utilizing the random Toeplitz hashing matrices, rather than evaluating the hash function directly on the raw data block, the function is applied to a seed beforehand to produce a structured pseudo-random sequence. Once raw data generation commences, the outcomes are processed online via an exclusive-OR operation with the prepared sequence. This streaming architecture thus reduces latency for real-time random number generation.

In this work, we extend the stream randomness extraction framework to the broad class of extractors based on (almost dual) universal$_2$ hash functions~\cite{carter1979universal,hayashi2016more}. We detail a general procedure for converting these extractors from a standard block-wise operation into a continuous streaming mode. Conditioned on any accessible quantum side information, given a reliable lower bound promise on the smooth min-entropy of the raw quantum measurement outcomes that are to come~\cite{renner2005universally,tomamichel2011leftover,portmann2022security}, this streaming approach securely yields nearly perfect random bits in full entropy online. To evaluate this framework, we present numerical studies on representative structured families, including the standard Toeplitz hashing construction~\cite{bennett1988privacy,ma2013postprocessing}, the circulant construction~\cite{foreman2025cryptomite}, and a modified Toeplitz hashing construction~\cite{hayashi2016more}, demonstrating how underlying algebraic structures influence both seed consumption and processing latency at a fixed security level.

\section{Preliminaries}
Before we commence, we review several notions that are essential for our results.

\subsection{Conditional min-entropy}
\begin{definition}
    Consider a quantum state $\rho_{XE}$. The min-entropy of system $X$ conditioned on system $E$ is defined as
    \begin{equation}
        H_{\mathrm{min}}(X|E)_\rho=\sup_{\sigma\in\mathcal{D}(\mathcal{H}_E)}\sup_{\lambda}\{\lambda\in\mathbb{R}:\rho_{XE}\leq e^{-\lambda}\mathbb{I}\otimes\sigma\}.
    \end{equation}
\end{definition}
Here, we use $\mathcal{D}(\mathcal{H})$ to represent the set of density operators acting on the Hilbert space $\mathcal{H}$.
In the context of randomness extraction, $\rho_{XE}$ corresponds to a classical-quantum state:
\begin{equation}
    \rho_{XE}=\sum_{x}p(x)\ket{x}\bra{x}\otimes\rho_E(x),
\end{equation}
with $\{p(x)\}$ a probability distribution. In this work, we use capital letters to represent random variables or quantum systems, and the lowercase letters denote a specific value of the corresponding random variable. As shown by the quantum left-over hash lemma, the min-entropy characterizes the amount of true randomness in the random variable $X$ conditioned on quantum side information $E$. Nevertheless, the conditional min-entropy is often too rigid and trivially equals zero. To bypass the problem, one can consider the smooth conditional min-entropy. The $\varepsilon$-smooth min-entropy of $X$ conditioned on $E$ is defined via the following optimization:
\begin{equation}
    H_{\mathrm{min}}^{\varepsilon}(X|E)_\rho=\max_{\substack{P(\rho',\rho)\leq\varepsilon \\\rho'\in\mathcal{D}(\mathcal{H}_{XE})}}H_{\mathrm{min}}(X|E)_{\rho'},
\end{equation}
where $P(\rho,\rho')$ denotes the purified distance between the two density operators, $\rho$ and $\rho'$.

\subsection{Randomness extractor}\label{sec:Ext}
If the min-entropy of a random variable $X$ conditioned on all possible quantum side information is well estimated, then there exists a seeded randomness extraction procedure that processes it into a nearly full-entropy random variable. Such a randomness extractor requires a seed. That is, given the random bit string to be processed, $X\in\{0,1\}^{n}$, with the guarantee that $H_{\mathrm{min}}(X|E)_\rho\geq k$, using a seed of perfectly random bits $Y\in\{0,1\}^d$, the randomness extractor function $\mathrm{Ext}$ processes it into some bit string $Z=\mathrm{Ext}(X,Y)\in\{0,1\}^m$ such that $H_{\mathrm{min}}^{\varepsilon}(Z|E)_\rho=m$ for some negligible $\varepsilon$. We denote such a randomness extractor a quantum-proof $(n,k,d,m,\varepsilon)$ extractor.

A notable fact is that there are randomness extractors that are strong in the random bits to be processed such that the seed can be re-used. To be rigorous,
\begin{definition}
    Consider an $(n,k,d,m,\varepsilon)$ quantum-proof randomness extractor function $\mathrm{Ext}:\{0,1\}^n\times\{0,1\}^d\rightarrow\{0,1\}^m$, if for all classical-quantum states $\rho_{XE}$ with conditional min-entropy $H_{\mathrm{min}}(X|E)_\rho\geq k$ and a perfect seed $Y$, we have
    \begin{equation}\label{eq:strongExt}
        \frac{1}{2}\|\rho_{\mathrm{Ext}(X,Y)YE}-\rho_{U_m}\otimes\rho_Y\otimes\rho_E\|\leq\varepsilon,
    \end{equation}
    where $\|\cdot\|$ represents the trace norm, $\rho_{U_m}=\mathbb{I}/2^m$ represents the fully mixed state given by a perfectly random bit string of $m$ bits, and $\rho_Y=\mathbb{I}/2^d$ represents the fully mixed state given by the seed $Y$, then we call the randomness extractor a strong extractor.
\end{definition}

Given a strong randomness extractor as above, if the $n$-bit string to be processed has smooth min-entropy $H_{\mathrm{min}}^{\varepsilon_s}(X|E)_\rho\geq k$, then the right hand side of Eq.~\eqref{eq:strongExt} is modified as $2\varepsilon_s+\varepsilon$.

\subsection{Constructions of (almost dual) universal$_2$ hashing functions}\label{sec:HashConstruct}
In this work, we focus on randomness extractors constructed from random (dual) universal$_2$ hashing functions.
Specifically, let $\mathrm{H}=\{h_y:\{0,1\}^n\rightarrow\{0,1\}^m\}_{y\in\mathcal{Y}}$ be a family of hash functions indexed by a random variable $Y$, which is the seed chosen uniformly at random and independently of the entropy source $X$.
The family $\mathrm{H}$ is \emph{universal$_2$} if for all $x\neq x'\in\{0,1\}^n$,
\begin{equation}
    \Pr_{y}\!\left[h_{y}(x)=h_{y}(x')\right]\le 2^{-m}.
    \label{eq:univ2}
\end{equation}
Universal$_2$ hashing functions provide a canonical construction of quantum-proof randomness extractors as per the quantum leftover hash lemma~\cite{carter1979universal,bennett1988privacy,bennett1995generalized}.
\begin{theorem}
    Given a family of universal$_2$ hash functions $\mathrm{H}=\{h_y:\{0,1\}^n\rightarrow\{0,1\}^m\}_{y\in\{0,1\}^d}$, it forms an $(n,k,d,m,2^{(m-k)/2})$ quantum-proof strong randomness extractor.
\end{theorem}

The conventional approach to constructing a family of universal$_2$ hash functions is through the Toeplitz matrices~\cite{krawczyk1994lfsr,krawczyk1995new}. An $m\times n$ Toeplitz matrix is in the form of
\begin{equation}\label{eq:Toeplitz}
    T_{y}=\begin{pmatrix}
        y_0 & y_{-1} &  \cdots & y_{-(n-1)} \\
        y_1 & y_0  & \cdots & y_{-(n-2)} \\
        \vdots &  \vdots & \ddots & \vdots \\
        y_{m-1} & y_{m-2}  & \cdots & y_{-n+m}
    \end{pmatrix},
\end{equation}
which is determined by the $(n+m-1)$-dimensional binary-valued random vector $y=(y_{-(n-1)},\cdots,y_{m-1})^{\mathrm{T}}$ over $\mathbb{F}_2$. For consistency throughout the work, when treating a random variable as a random vector, we shall denote it as a column vector by default, and $\mathrm{T}$ denotes the matrix transpose operation. This random vector is the seed random variable $Y$ in this extractor construction. All the $m\times n$ Toeplitz matrices comprise a family of $(n,k,d,m,2^{(m-k)/2})$ universal$_2$ hash functions with $d=m+n-1$. Upon a randomly chosen $Y=y\in\{0,1\}^{n+m-1}$, the raw data of an $n$-bit string $x$ is processed into as
\begin{equation}\label{eq:linearhash}
    z=h_{y}(x):= T_y\cdot x \bmod 2,
\end{equation}
where $\cdot$ is the standard matrix multiplication operation. When implementing the above calculation, one can apply the fast Fourier transform because of the highly symmetric structure of Toeplitz matrices. Append $x$ with $(m-1)$ zeros to construct a vector $x'$, i.e., $x'=(x\|0^{m-1})$, which is an $(n+m-1)$-bit string. For notation convenience, we use $(\cdot\|\cdot)$ to represent matrix (or vector) concatenation in a column, where the two concatenated matrices have the same number of columns, and we will use $(\cdot,\cdot)$ to represent matrix (or vector) concatenation in a row, where the two concatenated matrices have the same number of rows. Then,
\begin{equation}\label{eq:ToeplitzFFT}
    (c_{n-1}\|T_y\cdot x)=F^{-1}[F(y)\circ F(x')],
\end{equation}
where $c_{n-1}$ represents some $(n-1)$-dimensional binary vector, $F$ represents the discrete Fourier transform, $F^{-1}$ represents the inverse discrete Fourier transform, and $\circ$ represents the Hadamard product between vectors. The time complexity for this calculation is $O((n+m) \log (n+m))$. We provide the exact number of operations for its calculation in Appendix.

A drawback of the extractor construction from Toeplitz matrices is that the seed length is relatively long. To mitigate this, other constructions of universal$_2$ hash functions have been proposed. One such modification is the circulant hash function \cite{araujo2021circulant}. One such construction is the circulant function. For $n$ a prime with primitive root $2$, consider an $(n-1)$-bit raw data string $x=(x_0,\cdots,x_{n-2})^{\mathrm{T}}\in\{0,1\}^{n-1}$ and an $n$-bit seed $y=(y_0,\cdots,y_{n-1})^{\mathrm{T}}\in\{0,1\}^n$. Then the circulant construction goes as:
\begin{enumerate}
    \item Concatenate $x$ with a zero and extend it to an $n$-bit string, $x'=(x\|0)\in\{0,1\}^n$, with $x_{n-1}'=0$ and $x_i'=x_i$ for $0\leq i\leq n-2$.
    \item Generate a circulant matrix from $x'$:
    \begin{equation}\label{eq:CircMatrix}
    C_{x'}=\begin{pmatrix}
        x_0' & x_1' &  \cdots & x_{n-1}' \\
        x_{n-1}' & x_0'  & \cdots & x_{n-2}' \\
        \vdots &  \vdots & \ddots & \vdots \\
        x_1' & x_2'  & \cdots & x_0'
        \end{pmatrix}.
    \end{equation}
    \item Apply the matrix product between $C_{x'}$ and $y$ and preserve the first $m$ bits:
    \begin{equation}
        z=h_{y}(x):= (C_{x'}\cdot y \bmod 2)_0^{m-1},
    \end{equation}
    where we use $(\cdot)_{i}^{j}$ to denote the $i$'th to the $j$'th bits in the underlying vector.
\end{enumerate}
In this construction, the seed length is $1$ bit longer than the raw data. The random function defined by the circulant construction is universal$_2$ as in Eq.~\eqref{eq:univ2}. It is hence a quantum-proof strong extractor parameterized as $(n-1,k,n,m,2^{(m-k)/2})$. For a general $n$ that is not a prime with primitive root $2$, one can pad the raw data vector with zeros to round its size to the nearest prime with primitive root 2 and then apply the above construction. Similar to the Toeplitz hash construction, because of the highly symmetric structure of the circulant matrix, one can apply discrete Fourier transform to the matrix-vector multiplication. Specifically,
\begin{equation}\label{eq:CircFFT}
    (C_{x'}\cdot y)_i=\sum_{j=0}^{n-1}\mathrm{R}(x')_{i-j}\cdot y_j,
\end{equation}
where $\mathrm{R}(x')=(x_0,x_{n-1},x_{n-2},\cdots,x_1)$, and the indices $i,j$ and $i-j$ are $\mathrm{mod}\ n$. The overall computational time has complexity $O(n\log n)$.

The conventional Toeplitz hashing and the circulant constructions are based on universal$_2$ hash functions. In \citet{hayashi2016more}, it was observed and proven that one can apply the almost dual universal$_2$ hash functions for randomness extraction. Consider a surjective random function $f_y:\mathbb{F}_2^{n}\rightarrow\mathbb{F}_2^{m}$ with respect to the random variable $Y$. If it satisfies that $\forall x\in\mathbb{F}_2^{n}\backslash\{0\}^{n}$,
\begin{equation}
    \Pr_{y}\!\left[x\in(\ker{f_y})^\bot\right]\le \delta2^{-(n-m)},
\end{equation}
where $\ker{f_y}$ denotes the kernel space of the function $f_y$ and $\ker{f_y}^\bot$ denotes its orthogonal space, then we say $f_y$ is $\delta$-almost dual universal$_2$. Based on this notion, a modified Toeplitz hashing construction is proposed \cite{hayashi2016more}. Consider an $(n-1)$-bit random vector $y=(y_{1-m},\cdots,y_{n-m-1})^{\mathrm{T}}\in\{0,1\}^{n-1}$, let
\begin{equation}
    M_{y}=\begin{pmatrix}
        y_0 & y_{1} &  \cdots & y_{n-m-1} \\
        y_{-1} & y_0  & \cdots & y_{n-m-2} \\
        \vdots &  \vdots & \ddots & \vdots \\
        y_{1-m} & y_{2-m}  & \cdots & y_{n-2m}
    \end{pmatrix}.
\end{equation}
Notice that $M_y=T_y^{\mathrm{T}}$, where $T_y^{\mathrm{T}}$ represents the $(n-m)\times m$ Toeplitz matrix constructed from $y$ as in Eq.~\eqref{eq:Toeplitz}. Then one can apply the following function:
\begin{equation}
    z=h_{y}(x):= (M_y,\mathbb{I}_m)\cdot x,
\end{equation}
to extract randomness from the $n$-bit string of raw data $x$. It has been proven that this construction constitutes an $(n,k,n-1,m,2^{(m-k)/2})$ randomness extractor. The above extraction function can also be calculated efficiently via the discrete Fourier transform, noticing that
\begin{equation}\label{eq:Hayashi}
    (M_y,\mathbb{I}_m)\cdot x=M_y\cdot\underline{x}\oplus\bar{x},
\end{equation}
with $\underline{x}=(x_0,\cdots,x_{n-m-1})\in\{0,1\}^{n-m}$ and $\bar{x}=(x_{n-m},\cdots,x_{n-1})\in\{0,1\}^{m}$, and that $M_y\cdot\underline{x}$ corresponds to the multiplication between a Toeplitz matrix and a vector as in Eq.~\eqref{eq:linearhash}. The matrix-vector product has a time complexity of $O(n\log n)$, and the bit-wise addition has a time complexity of $O(m)$.

\section{Theoretical Framework}\label{sec:theory}

Now, we demonstrate how to convert the randomness extractor constructions in Sec.~\ref{sec:HashConstruct} into stream versions. As a warm-up, we review the construction in Ref.~\cite{huang2022stream} that converts the conventional Toeplitz hash matrix construction. Consider a raw data bit string $x\in\{0,1\}^{n}$ with conditional min-entropy of $H(X|E)\geq k=m-2\log{\varepsilon}$, with $\varepsilon$ the tolerable soundness parameter. The random Toeplitz hashing using Toeplitz matrices of size $m\times n$ in Eq.~\eqref{eq:linearhash} shall process $x$ into an $m$-bit nearly perfect random bit string. In the stream version, one shall prepare a seed of perfectly random bits, $r\in\{0,1\}^{n-m}$, and use another seed $y\in\{0,1\}^{2n-m-1}$ to randomly choose a Toeplitz matrix $T_{y}$ of size $(n-m)\times n$ and calculate an $n$-bit string $w$ as
\begin{equation}\label{eq:mask}
    w=T_{y}^{\mathrm{T}}\cdot r,
\end{equation}
which we call a mask variable. Then, we can obtain nearly perfect random bits from
\begin{equation}\label{eq:stream}
    z=(x\oplus w)_0^{m-1},
\end{equation}
which are the first $m$ bits of the bit-wise addition of $x$ and $w$. This process can be seen as an extractor function of
\begin{equation}
    \mathrm{StreamT}:\{0,1\}^n\times\{0,1\}^{3n-2m-1}\rightarrow\{0,1\}^{m},
\end{equation}
which takes $(r\|y)$ as an $(3n-2m-1)$-bit seed and processes the $n$-bit raw data $x$ into an $m$-bit final data $z$. Ref.~\cite{huang2022stream} proves that this extractor has the same soundness parameter $\varepsilon$ as in its corresponding block version. Furthermore, the stream version of Toeplitz hash construction is strong in $(R\|Y)$ in the sense that the same seed variable $Y$ can be re-used to randomly draw Toeplitz matrices, and the last $(n-m)$ bits of $x\oplus w$ can be collected and used as the seed $R$ for another randomness extraction, albeit an increase of $\varepsilon$ in the soundness parameter.

Similarly, we can convert the circulant construction and the modified Toeplitz construction into stream versions. We first describe the protocols before explaining why they are valid. For the circulant construction, first prepare an $(n-m-1)$-bit seed $r$ and append it with $(m+1)$ zeros to obtain an $n$-bit string, $r'=(r\|0^{m+1})$. Then, as in Eq.~\eqref{eq:CircMatrix}, construct an $n\times n$ circulant matrix $C_{r'}$ from the vector $r'$. Next, use another $n$-bit seed $y$ and calculate an $(n-1)$-bit string $w$ as
\begin{equation}
    w = (C_{r'}\cdot y)_0^{n-2},
\end{equation}
which gives a mask variable $W$. Afterwards, one applies the process as in Eq.~\eqref{eq:stream} to process the raw data $x\in\{0,1\}^{n-1}$ and obtain the final output $z$. The whole process can be seen as an extractor function of
\begin{equation}
    \mathrm{StreamC}:\{0,1\}^{n-1}\times\{0,1\}^{2n-m-1}\rightarrow\{0,1\}^{m},
\end{equation}
which takes $(r\|y)$ as an $(2n-m-1)$-bit seed. The extractor has the same soundness parameter $\varepsilon$ as in its corresponding block version. Similarly to the stream version of the Toeplitz hash construction, the stream version of the circulant construction is strong in $(R\|Y)$.

For the stream version of the modified Toeplitz construction, first prepare an $(n-m)$-bit seed $r$. Next, use an $(n-1)$-bit seed $y=(y_{1+m-n},\cdots ,y_{m-1})^{\mathrm{T}}$ and construct
\begin{equation}
    M_{y}=\begin{pmatrix}
        y_0 & y_{1} &  \cdots & y_{m-1} \\
        y_{-1} & y_0  & \cdots & y_{m-2} \\
        \vdots &  \vdots & \ddots & \vdots \\
        y_{1+(m-n)} & y_{2+(m-n)}  & \cdots & y_{2m-n}
    \end{pmatrix},
\end{equation}
with which one calculates an $n$-bit string $w$ as
\begin{equation}\label{eq:ModifiedMask}
    w=(M_y,\mathbb{I}_{n-m})^{\mathrm{T}}\cdot r
\end{equation}
as a mask variable $W$. Afterwards, one applies the process as in Eq.~\eqref{eq:stream} to process the raw data $x$ and obtain the final output $z$. The whole process can be seen as an extractor function of
\begin{equation}
    \mathrm{StreamM}:\{0,1\}^n\times\{0,1\}^{2n-m-1}\rightarrow\{0,1\}^{m},
\end{equation}
which takes $(r\|y)$ as a $(2n-k-1)$-bit seed. The extractor has the same soundness parameter $\varepsilon$ as in its corresponding block version and strong in $(R\|Y)$.

We briefly explain why the stream versions of the randomness extraction work; a complete proof will be presented elsewhere. \citet{huang2022stream} has proven the validity of the stream version of the standard Toeplitz hashing randomness extractor, and the proof is based on the picture of \emph{error correction with quantum side information} \cite{tsurumaru2022equivalence}. In this primitive, consider that an $m$-bit message $t\in\{0,1\}^m$ is encoded into an $n$-bit codeword $c(t)\in\{0,1\}^n$, which is then sent through a noisy quantum channel and gives an $n$-bit output $x$. After considering the quantum side information introduced by the noisy quantum channel, the output $x$ is in the following state correlated with the noise:
\[
\rho(t)=\sum_{x\in\{0,1\}^n}\ket{x}\bra{x}\otimes\rho({c(t),x}).
\]
Then, a proper decoding process is applied to $x$ to recover the original message. For a successful transmission of the message, one can apply the random universal$_2$ hashing to construct an $[n,m]$ linear code. The standard construction is based on the standard Toeplitz matrices as described in Eq.~\eqref{eq:Toeplitz}.
\citet{tsurumaru2022equivalence} shows that the error correction with quantum side information is equivalent to the primitive of \emph{privacy amplification against quantum side information} (see \cite[Theorem 2]{tsurumaru2022equivalence} for the exact statement). In this primitive, an $n$-bit string classical message $x\in\{0,1\}^n$ is correlated with quantum side information:
\[
\rho_{XE}=\sum_{x\in\{0,1\}^n}\ket{x}\bra{x}\otimes\rho_E(x),
\]
which satisfies $H_{\mathrm{min}}(X|E)_{\rho}\geq k$. The privacy amplification takes a linear function and processes $x$ into an $m$-bit string $z\in\{0,1\}^m$ such that the final state is close to the following ideal state:
\[
\tilde{\rho}_{ZE}=\sum_{z\in\{0,1\}^m}\ket{z}\bra{z}\otimes\tilde{\rho}_E,
\]
where the quantum side information system $E$ becomes independent of $z$. \citet{hayashi2016more} shows that both the standard (almost) universal$_2$ hashing and the (almost) dual universal$_2$ hashing functions are sufficient for this task. Note that this primitive is exactly the task of quantum randomness extraction as we describe in Sec.~\ref{sec:Ext} \footnote{\citet{ma2013postprocessing,huang2022stream} implicitly prove the equivalence between the primitives of \emph{error correction with quantum side information} and \emph{privacy amplification against quantum side information} using the random Toeplitz hashing matrices.}. Combining the results of \citet{huang2022stream,tsurumaru2022equivalence,hayashi2016more}, one can show that it is valid to apply a similar approach as in \citet{huang2022stream} to the other block-wise randomness extraction constructions based on the (almost dual) universal$_2$ hashing functions, which transform them into a corresponding stream version.

\section{Numerical Results}\label{sec:numerical_simulation}
We implemented both the block and stream versions of the aforementioned randomness extraction constructions in MATLAB. To evaluate their practical performance, particularly computational cost measured by runtime, we apply these extractors to realistic quantum measurement data. Both the raw input data and the extraction seeds are sourced from the experimental datasets provided by \citet{nie2015generation}, which we pre-convert into bitstream files. Specifically, we utilize the \texttt{rawdata10G} file as our raw input, which is reported to have a min-entropy rate of at least $0.86$, and we utilize the \texttt{randomdata5G} file to supply the random seeds. In our simulation, for each input length \(n\), we fix an entropy-rate parameter $\alpha = k/n<0.86$.
For the runtime benchmark and the empirical statistical tests, we set the nominal output length to $m=k$.
The additive security overhead, e.g., the term $2\log_2(1/\varepsilon)$ in the leftover-hashing bound, is omitted in the timing comparison because it only changes the retained output length by an additive amount and does not affect the dominant FFT-based runtime scaling.
We consider two representative entropy rates:
\begin{equation}
    \alpha=0.5
    \quad\mathrm{and}\quad
    \alpha=0.8 .
\end{equation}
Choosing a conservative value of $\alpha$ below the estimated entropy rate does not compromise extraction security; it only reduces the nominal output rate and may increase the relative computational cost. 

For the block implementation of our numerical experiments, we define $t^{\rm block}_{(A)}(n,\alpha)$ as the median runtime required to read an $n$-bit raw data block from \texttt{rawdata10G} and perform a single block-wise randomness extraction, where $A\in\{\mathrm{Toeplitz},\mathrm{Circulant},\mathrm{Modified\ Toeplitz}\}$ denotes the chosen extractor construction. On the other hand, the stream implementation separates the computation into an offline preprocessing stage to prepare the mask variable, and a real-time processing stage that reads the raw data on the fly and applies a bitwise XOR with the mask. Accordingly, we define $t^{\rm mask}_{(A)}(n,\alpha)$ as the time required to generate the $n$-bit pseudo-random mask, and $t^{\rm xor}(n)$ as the time needed to read an $n$-bit raw block from \texttt{rawdata10G} and execute the bitwise XOR operation. The total runtime of the stream randomness extraction is then defined as
\begin{equation}
    t^{\rm stream}_{(A)}(n,\alpha)
    =
    t^{\rm mask}_{(A)}(n,\alpha)
    +
    t^{\rm xor}(n).
\end{equation}
In the figures below, the dashed stream-total curves show \(t^{\rm stream}_{(A)}\), while the open markers indicate \(t^{\rm mask}_{(A)}\) alone.

All numerical experiments are executed in MATLAB on a single computational thread. For each $(n,\alpha)$ setting, we initially perform five warm-up runs, which execute the standard read-and-compute routine without recording the runtime, to mitigate MATLAB's just-in-time compilation overhead, FFT initialization, and transient file-buffer effects. Subsequently, we measure the runtime across $N_{\rm rep}=100$ trials, reporting the median value as the representative runtime and recording the standard deviation to quantify timing fluctuations.

The computational bottleneck for all three extractor families and both the block and the stream implementations is the FFT-based convolution. For Toeplitz hashing, this manifests as a linear convolution of length $O(n+m)$ in the block implementation and $O(2n-m)$ for the stream mask generation. The circulant construction relies on a convolution of length $O(n)$, which holds for both the block and the stream implementations. In the modified Toeplitz construction, the primary operation is a Toeplitz-like convolution of length $O(n)$, which holds for both the block and the stream implementations. In the stream implementations for all the extractor families, the subsequent XOR steps scale linearly as simple bitwise operations. We provide a theoretical analysis of the overall computation costs for the implementations in Appendix A.

\begin{table}[tbp]
\caption{
Dominant vector lengths in the convolution computation. The computational bottleneck in each randomness extractor is the matrix-vector multiplication. In each implementation, this multiplication is reduced to an FFT-based convolution after the hashing matrices and raw data are represented by suitable vectors. The table lists their lengths.
}
\begin{ruledtabular}
\begin{tabular}{lcc}
Family & Block mode & Stream mode \\
\hline
Toeplitz & \(n+m-1\) & \(2n-m-1\) \\
Circulant & \(n+1\) & \(n+1\) \\
Modified Toeplitz & \(n-1\) & \(n-1\) \\
\end{tabular}
\end{ruledtabular}
\label{tab:runtime_kernels}
\end{table}

\begin{figure}[tbp]
    \centering
    \includegraphics[width=0.98\columnwidth]{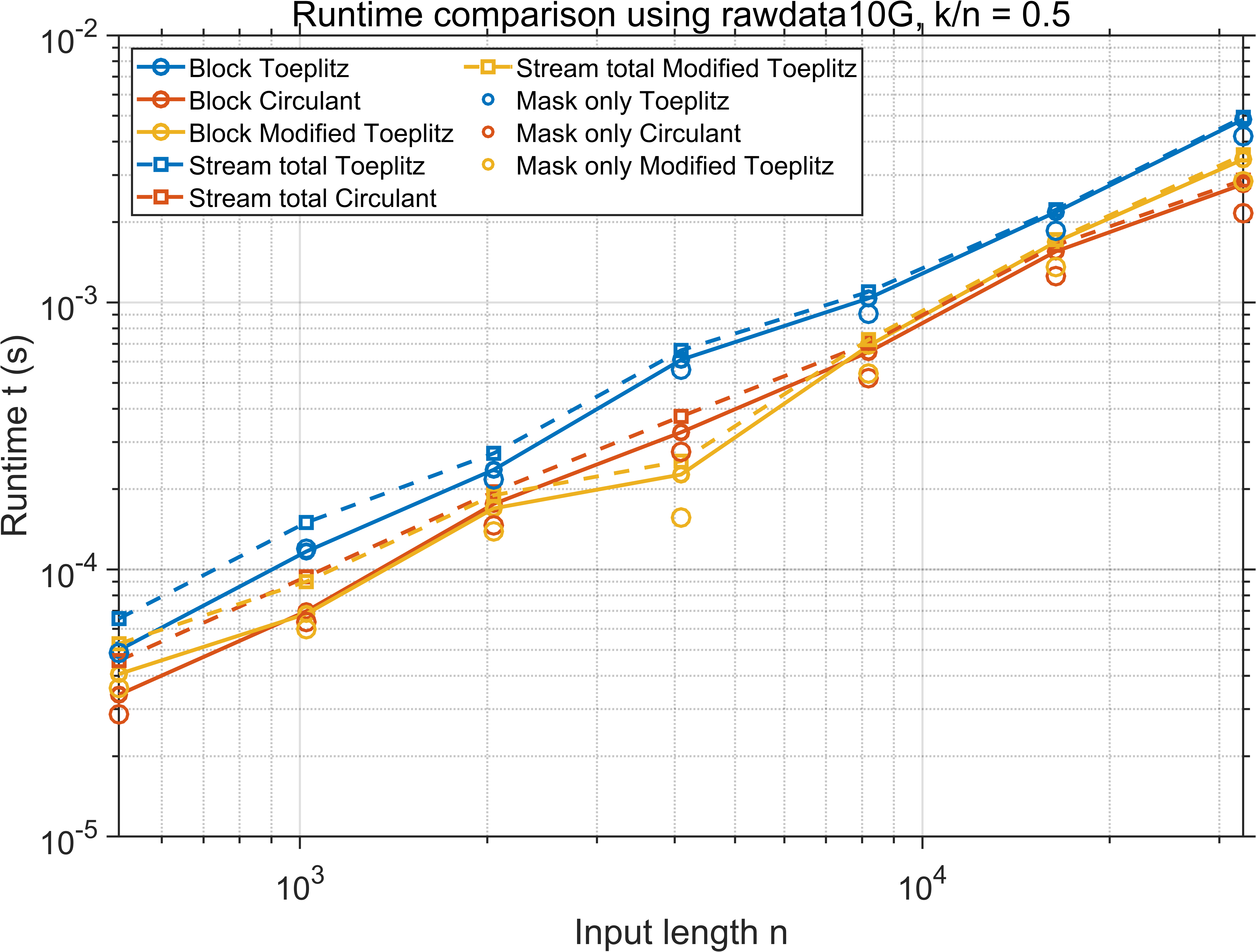}
    \caption{
    Runtime comparison between block and stream implementations at entropy rate \(k/n=0.5\).
    The solid curves show \(t_A^{\rm block}\) for $A\in\{\mathrm{Toeplitz},\mathrm{Circulant},\mathrm{Modified\ Toeplitz}\}$, the runtime of one block-wise hash evaluation after reading an \(n\)-bit raw block from \texttt{rawdata10G}.
    The dashed curves show the total stream runtime \(t_A^{\rm stream}=t_A^{\rm mask}+t_A^{\rm xor}\).
    Open markers on the stream curves indicate the mask-generation time \(t_A^{\rm mask}\) alone.
    }
    \label{fig:runtime_alpha50}
\end{figure}

\begin{figure}[tbp]
    \centering
    \includegraphics[width=0.98\columnwidth]{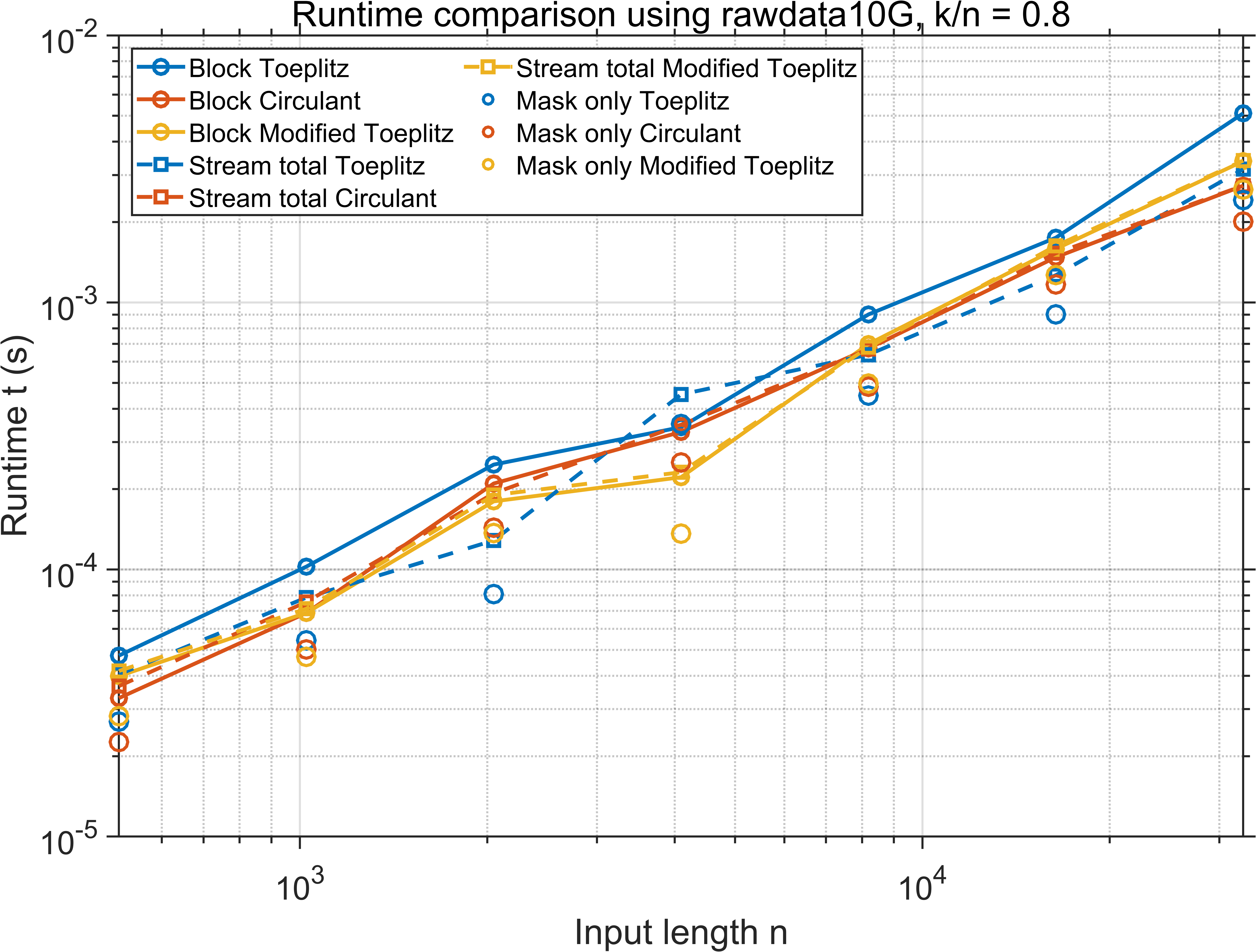}
    \caption{
    Runtime comparison between block and stream implementations at entropy rate \(k/n=0.8\).
    The same benchmarking protocol as in Fig.~\ref{fig:runtime_alpha50} is used.
    At this higher entropy rate, the Toeplitz stream implementation benefits from a shorter effective convolution length in the offline mask-generation stage, while the circulant and modified Toeplitz constructions remain mainly governed by FFT lengths set by \(n\).
    }
    \label{fig:runtime_alpha80}
\end{figure}

Figures~\ref{fig:runtime_alpha50} and~\ref{fig:runtime_alpha80} present the results of our numerical experiments. Focusing on the runtime comparison at an entropy rate of $k/n=0.5$ (Figure~\ref{fig:runtime_alpha50}), the standard Toeplitz construction exhibits the highest runtime among the three families, whereas the circulant construction achieves the lowest runtime for large $n$. Additionally, the stream-total curves plot slightly above their corresponding mask-only points, reflecting the added overhead of the online read-and-XOR stage included in $t_{\rm stream}$.
Figure~\ref{fig:runtime_alpha80} displays the corresponding runtime comparison at an entropy rate of $k/n=0.8$. At this higher rate, the Toeplitz stream implementation becomes noticeably faster than its block counterpart. This aligns with theoretical expectations based on effective convolution lengths: the block hash scales with a length of $O(n+m)$, whereas the stream mask generation scales with $O(2n-m)$, which is significantly shorter when $k/n=0.8$. In contrast, the dominant FFT lengths for the circulant and modified Toeplitz constructions depend primarily on $n$; thus, their block and stream-total runtimes remain comparable once the read-and-XOR stage is accounted for.

Overall, these numerical results support the main operational picture of stream extraction: the nontrivial hash computation is shifted to an offline mask-generation step, while the online stage is reduced to reading the raw block and performing a simple XOR operation. 
The relative advantage of the stream implementation depends on the entropy rate and on the algebraic structure of the hash family.

\paragraph*{Randomness test.}

As an additional empirical sanity check, we apply the NIST SP~800-22 statistical test suite to the random bits generated by the three stream implementations.
As analyzed in Sec.~\ref{sec:theory}, each stream extractor retains the first $m$ bits of the $n$-bit string $x\oplus w$ as its final output, denoted as $z=(x\oplus w)_0^{m-1}$.
The NIST statistical tests are performed on sequences of $10^6$ bits from $z$. 
For an entropy rate of \(k/n=0.5\), we evaluate \(65\) sequences, yielding \(6.5\times10^7\) tested bits per extractor. 
For \(k/n=0.8\), we evaluate \(104\) sequences, totaling \(1.04\times10^8\) tested bits per extractor. 
All six extracted bitstreams successfully pass the applicable NIST tests under these partitions. 
A summary of these results is provided in Table~\ref{tab:nist_summary}, with the detailed testing procedure outlined in Appendix~\ref{app:nist_details}.

We emphasize that the NIST tests serve merely as empirical validations of the generated finite bitstreams. 
They do not constitute a security proof; the security of the extracted outputs is fundamentally guaranteed by the quantum universal hash lemma as discussed in Sec.~\ref{sec:theory}.

\begin{table}[tbp]
\caption{
Summary of the NIST SP~800-22 statistical sanity check.
The NIST tests are used only as finite-sample statistical sanity checks.
}
\begin{ruledtabular}
\begin{tabular}{lccc}
Extractor & \(k/n\) & Tested bits & Result \\
\hline
Toeplitz & 0.5 & \(6.5\times10^7\) & Pass \\
Toeplitz & 0.8 & \(1.04\times10^8\) & Pass \\
Circulant & 0.5 & \(6.5\times10^7\) & Pass \\
Circulant & 0.8 & \(1.04\times10^8\) & Pass \\
Modified Toeplitz & 0.5 & \(6.5\times10^7\) & Pass \\
Modified Toeplitz & 0.8 & \(1.04\times10^8\) & Pass \\
\end{tabular}
\end{ruledtabular}
\label{tab:nist_summary}
\end{table}

\section{Conclusions}

We have developed a general framework for converting block-wise randomness extractors based on the random (almost dual) universal$_2$ hashing constructions into streaming extractors. 
The construction separates the extraction procedure into an offline mask-generation stage, which is prior to data accumulation, and an online read-and-XOR stage, which is executed in real time as the raw data from quantum measurements is generated. 
Therefore, our method shifts the time-consuming hash computation from post-acquisition processing to offline preprocessing and removes it from the real-time data-acquisition path. 
Within the quantum-proof leftover-hashing framework, the resulting stream extractors preserve the composable security guarantees of their block counterparts against quantum side information, provided that the raw data to be accumulated is promised with a proper lower bound on its (smooth) quantum min-entropy.

We applied our framework to three representative structured hashing families: standard Toeplitz hashing, circulant hashing, and the modified Toeplitz construction. 
These examples illustrate how the algebraic structure of the hash family determines the seed description length, the effective FFT kernel, and the runtime advantage of the stream implementation. 
In particular, the Toeplitz stream construction can substantially reduce the effective convolution length at high entropy rate, while the circulant and modified Toeplitz constructions provide shorter structured descriptions and comparable FFT-dominated scaling.

Our numerical benchmarks, performed on pre-converted raw bitstreams, confirm the expected separation between offline and online costs. 
The stream-total runtime is well described by the sum of the mask-generation time and the read-and-XOR time, and the online stage remains a lightweight bitwise operation. 
The generated extracted bitstreams also pass the NIST SP~800-22 statistical tests, serving as an empirical sanity check in addition to the information-theoretic security analysis.

The present framework provides a practical route toward low-latency post-processing for QRNGs and related quantum-information protocols where raw randomness is produced continuously or slowly. 
It also clarifies the design space of streaming extractors: different universal-hashing families trade seed length, preprocessing cost, buffering requirement, and online latency in distinct ways. 
Future work may extend the same approach to broader almost-universal hashing families, hardware-level implementations, and adaptive extraction protocols for time-varying entropy sources.


\begin{acknowledgments}
The authors thank colleagues and collaborators for helpful discussions on streaming randomness extraction, universal hashing, and numerical benchmarking. 
We also thank the members of the related quantum information and quantum random number generation communities for valuable comments on the manuscript.
\end{acknowledgments}

\section*{Data Availability Statement}

The MATLAB scripts, processed runtime data, extracted bitstreams used as inputs to the NIST SP~800-22 statistical tests, and the corresponding NIST final analysis reports are available at \url{https://doi.org/10.5281/zenodo.20542218}. 
The original quantum random number generation data used in the numerical benchmark are from Ref.~\cite{nie2015generation}. 
Additional information is available from the corresponding authors upon reasonable request.

\section*{Funding}

No external funding was received for this work.

\section*{Conflict of Interest}

The authors declare no conflict of interest.

\appendix

\section{Theoretical analysis of computational complexity}
\label{app:complexity}

We provide a basic operation-count estimate for the three structured hash families considered in the main text.
The purpose of this appendix is to complement the empirical MATLAB runtimes by giving the dominant FFT-based scaling of the corresponding block-mode computations.
Here a ``basic operation'' refers to a scalar addition or multiplication, and the logarithm in the FFT length is understood as base two.

\paragraph*{Standard Toeplitz hashing.}

We first count the number of basic operations in the standard Toeplitz hashing construction.
For an \(m\times n\) Toeplitz matrix, the FFT-based multiplication embeds the Toeplitz product into a linear convolution of length \(n+m-1\).
Let
\begin{equation}
    K_{\rm T}=2^{\lceil \log_2(n+m-1)\rceil}
\end{equation}
be the padded FFT length. We remark that padding vectors to this length, which is a power of two, is necessary for a variety of FFT algorithms.
Appending \(x\) with \(m-1\) zeros requires \(m-1\) padding operations, and extending both convolution vectors to length \(K_{\rm T}\) requires \(2[K_{\rm T}-(n+m-1)]\) additional zero-padding operations.
The two forward FFTs and one inverse FFT require approximately \(3K_{\rm T}\log_2 K_{\rm T}\) operations, and the pointwise Hadamard product requires \(K_{\rm T}\) operations.
Thus, up to implementation-dependent constants, the block-mode Toeplitz construction requires
\begin{equation}
\begin{aligned}
    \tilde{t}_{\mathrm{Toeplitz}}^{\mathrm{block}}
    =&
    3K_{\rm T}\log_2 K_{\rm T}
    +K_{\rm T}
    +(m-1) \\
    &+2[K_{\rm T}-(n+m-1)]  \\
    \approx&
    3K_{\rm T}(\log_2 K_{\rm T}+1)-2n-m.
\end{aligned}
\end{equation}
We denote the theoretical time complexity with $\tilde{t}$ to distinguish it from the realistic computational time $t$ in the main text.
This gives the expected \(O((n+m)\log(n+m))\) scaling. Here and in later discussions, we neglect small constants, which are negligible compared to the other terms that scale with respect to $n$ or $m$.

\paragraph*{Stream Toeplitz hashing.}

In the stream version of a randomness extractor construction, the number of basic operations in executing the exclusive-OR between the raw data and the mask is $n$; this is the same for all the constructions in this work. The main computational cost lies in the preparation of the mask. In the case of the standard Toeplitz hashing, one starts with an $(n-m)$-bit seed $r$ and applies a Toeplitz matrix of size $(n-m)\times n$ to it. The Toeplitz matrix is determined by a random vector of $(2n-m-1)$ bits. To apply the FFT-based multiplication, we shall pad zeros to this random vector and $r$ respectively and transform them into vectors of a length
\begin{equation}
    K_{\mathrm{T}}'=2^{\lceil\log_2(2n-m-1)\rceil}.
\end{equation}
This step requires $2K_{\mathrm{T}}'-3n+2m$ zero-padding operations. The FFT-based matrix-vector multiplication then takes approximately $3K_{\mathrm{T}}'\log_2K_{\mathrm{T}}'+K_{\mathrm{T}}'$ basic operations. In all, to prepare for the mask in the stream version of the standard Toeplitz hashing construction, the number of basic operations is
\begin{equation}
\begin{split}
    \tilde{t}_{\mathrm{Toeplitz}}^{\mathrm{stream}}\approx 3K_{\mathrm{T}}'(\log_2K_{\mathrm{T}}'+1)-3n+2m.
\end{split}
\end{equation}

\paragraph*{Circulant hashing.}

For the circulant construction, consider the ideal case in which the circulant dimension satisfies the required number-theoretic condition.
In the runtime benchmark, we use a common raw input length \(n\) and an associated circulant dimension \(n+1\).
The matrix-vector product is evaluated as a cyclic convolution.
Let
\begin{equation}
    K_{\rm C}=2^{\lceil \log_2(n+1)\rceil}
\end{equation}
be the padded FFT length.
The permutation \(R(x)=(x_0,x_{n},x_{n-1},\ldots,x_1)\) is a fixed reordering and contributes only a lower-order cost.
The dominant cost comes from two forward FFTs, one inverse FFT, and one Hadamard product.
Therefore, the block-mode circulant construction requires approximately
\begin{equation}
    \tilde{t}_{\mathrm{Circulant}}^{\mathrm{block}}
    \approx
    3K_{\rm C}(\log_2 K_{\rm C}+1)-2n,
\end{equation}
up to lower-order padding and permutation terms.
The scaling is \(O(n\log n)\).

\paragraph*{Stream circulant hashing.}

In the stream version of the circulant hashing, one starts from an $(n-m-1)$-bit seed $r$ and appends it with $(m+1)$ zeros. Let
\begin{equation}
    K_{\mathrm{C}}'=2^{\lceil\log_2{n}\rceil},
\end{equation}
and pad zeros to $r'$ and the $n$-bit random vector determining the circulant matrix. In all, the zero padding operation takes approximately $2K_{\mathrm{C}}'-2n+m$ basic operations. The FFT-based matrix-vector multiplication takes approximately $3K_{\mathrm{C}}'\log_2{K_{\mathrm{C}}'}+K_{\mathrm{C}}'$ basic operations. Therefore, to prepare for the mask in the stream version of the circulant hashing construction, the number of basic operations is 
\begin{equation}
    \tilde{t}_{\mathrm{Circulant}}^{\mathrm{stream}}\approx 3K_{\mathrm{C}}'(\log_2{K_{\mathrm{C}}'}+1)-2n+m.
\end{equation}

\paragraph*{Modified Toeplitz hashing.}

In the modified Toeplitz construction, the computation decomposes into a Toeplitz-like matrix-vector product and a bitwise XOR with the identity part.
The nontrivial Toeplitz-like component has effective convolution length \(n-1\).
Let
\begin{equation}
    K_{\rm M}=2^{\lceil \log_2(n-1)\rceil}.
\end{equation}
The FFT-based Toeplitz-like multiplication contributes approximately
\[
    3K_{\rm M}(\log_2 K_{\rm M}+1)-2n+m
\]
basic operations.
The final XOR with the identity part contributes \(m\) additional bitwise operations.
Hence, the total block-mode operation count is approximately
\begin{equation}
    \tilde{t}_{\mathrm{Modified}}^{\mathrm{block}}
    =
    3K_{\rm M}(\log_2 K_{\rm M}+1)-2n+2m.
\end{equation}
The dominant scaling is again \(O(n\log n)\), with the non-FFT identity and XOR operations contributing only linear-time overhead.

\paragraph*{Stream modified Toeplitz hashing.}

In the stream version of the modified Toeplitz hashing, the matrix-vector multiplication to prepare the mask variable $w$ is essentially the calculation of $M_y^{\rm T}\cdot r$, where $M_y$ is a Toeplitz matrix of size $(n-m)\times m$, and $r$ is a random seed of $(n-m)$ bits. For this process, similar to the above analyses, one will apply approximately $2K_{\rm M}'-2n+m$ zero-padding operations, with
\begin{equation}
    K_{\rm M}'=2^{\lceil\log_2(n-1)\rceil}.
\end{equation}
The FFT-based matrix-vector multiplication takes $3K_{\rm M}'\log_2{K_{\rm M}'}+K_{\rm M}'$ basic operations. Note that in Eq.~\eqref{eq:ModifiedMask}, one also needs to apply a vector concatenation that appends $r$ to $M_{y}^{\rm T}\cdot r$, which takes $(n-m)$ basic operations. In all, to prepare for the mask in the stream version of the modified Toeplitz hashing construction, the number of basic operations is 
\begin{equation}
    \tilde{t}_{\mathrm{Modified}}^{\mathrm{stream}}\approx 3K_{\mathrm{M}}'(\log_2{K_{\mathrm{M}}'}+1)-n.
\end{equation}

In summary, we list the asymptotic complexity scaling of all the methods in Table~\ref{tab:TimeComplexity}. For the stream constructions, we list both their total time complexity including the preparation of the mask and the exclusive-OR operation. We note that the circulant construction requires $n$ to be a prime with primitive root $2$, and a padding operation needs to be applied for a general $n$. This process is not included in the table.

\begin{table*}[tbp]
\caption{Asymptotic time complexity for different randomness extraction constructions.}
\centering
\begin{ruledtabular}
    \begin{tabular}{lcc}
    Mode & Construction & Time complexity \\
    \hline
    \multirow{3}{*}{Block} & Standard Toeplitz & $O(3(n+m)\log(n+m)+n+2m)$ \\
     & Circulant & $O(3n\log{n}+n+m)$ \\
     & Modified Toeplitz & $O(3n\log{n}+n+2m)$ \\
    \hline
    \multirow{3}{*}{Stream} & Standard Toeplitz & $O(3(2n-m)\log(2n-m)+4n-m)$ \\
     & Circulant & $O(3n\log{n}+2n+m)$ \\
     & Modified Toeplitz & $O(3n\log{n}+3n)$ \\
    \end{tabular}
\end{ruledtabular}
\label{tab:TimeComplexity}
\end{table*}

\section{Details of the NIST statistical sanity check}
\label{app:nist_details}

This appendix describes the preparation of the input bitstreams and the sequence partition used in the NIST SP~800-22 statistical sanity check.
The tests are applied to the final extracted outputs of the stream implementations, rather than to the raw data, the mask variables, or the complete \(n\)-bit masked strings.

In the numerical implementation used for the NIST test, the raw bitstream is partitioned into \(N_{\rm blk}=102400\) blocks, each containing
\begin{equation}
    n=1280
\end{equation}
raw bits.
Thus, the total amount of raw data used to generate the extracted bitstreams is
\begin{equation}
    N_{\rm blk} n
    =
    102400\times1280
    =
    1.31072\times10^8
\end{equation}
bits.
The block length \(n=1280\) is the numerical block size used in the MATLAB implementation and runtime benchmark; it is not a restriction of the theoretical stream-extraction framework.

For each block, the stream implementation first generates an \(n\)-bit mask \(w\), computes the masked string \(x\oplus w\), and then retains only the first \(m=k\) bits,
\begin{equation}
    z=(x\oplus w)_0^{m-1}.
\end{equation}
The concatenation of these extracted outputs is used as the input to the NIST statistical test suite.

For \(k/n=0.5\), one has \(k=m=640\), so each generated extracted bitstream contains
\begin{equation}
    N_{\rm blk}\times k
    =
    102400\times640
    =
    6.5536\times 10^7
\end{equation}
bits.
In the NIST test, we use \(65\) sequences of length \(10^6\) bits, corresponding to \(6.5\times10^7\) tested bits.
The remaining \(5.36\times10^5\) bits are not used because the NIST test is run with complete \(10^6\)-bit sequences.

For \(k/n=0.8\), one has \(k=m=1024\), so each generated extracted bitstream contains
\begin{equation}
    N_{\rm blk}\times k
    =
    102400\times1024
    =
    1.048576\times 10^8
\end{equation}
bits.
In the NIST test, we use \(104\) sequences of length \(10^6\) bits, corresponding to \(1.04\times10^8\) tested bits.
The remaining \(8.576\times10^5\) bits are not used because the NIST test is run with complete \(10^6\)-bit sequences.

The repository files listed in Table~\ref{tab:nist_files} correspond to the extracted-output bitstreams used as inputs to the NIST test and archived in the Zenodo repository cited in the Data Availability Statement.
All six extracted bitstreams pass the applicable NIST SP~800-22 statistical tests under the tested sequence partitions.

\begin{table*}[tbp]
\caption{
Input bitstreams and sequence partitions used for the NIST SP~800-22 statistical sanity check.
The repository files correspond to the extracted-output bitstreams archived in the Zenodo repository cited in the Data Availability Statement.
}
\begin{ruledtabular}
\begin{tabular}{lccccc}
Extractor & \(k/n\) & Repository file & Sequence length & Number of sequences & Result \\
\hline
Toeplitz & 0.5 & \texttt{NIST\_FinalRD\_Toeplitz\_alpha50.bin} & \(10^6\) & 65 & Pass \\
Circulant & 0.5 & \texttt{NIST\_FinalRD\_Circulant\_alpha50.bin} & \(10^6\) & 65 & Pass \\
Modified Toeplitz & 0.5 & \texttt{NIST\_FinalRD\_ModifiedToeplitz\_alpha50.bin} & \(10^6\) & 65 & Pass \\
Toeplitz & 0.8 & \texttt{NIST\_FinalRD\_Toeplitz\_alpha80.bin} & \(10^6\) & 104 & Pass \\
Circulant & 0.8 & \texttt{NIST\_FinalRD\_Circulant\_alpha80.bin} & \(10^6\) & 104 & Pass \\
Modified Toeplitz & 0.8 & \texttt{NIST\_FinalRD\_ModifiedToeplitz\_alpha80.bin} & \(10^6\) & 104 & Pass \\
\end{tabular}
\end{ruledtabular}
\label{tab:nist_files}
\end{table*}

The NIST tests are not interpreted as a proof of security.
They are used only to verify that the finite extracted bitstreams do not exhibit obvious statistical defects under standard empirical tests.
The composable security of the extracted outputs is instead guaranteed by the quantum-proof universal-hashing framework discussed in the main text.

\bibliography{Refs}

@article{portmann2022security,
  title        = {Security in Quantum Cryptography},
  author       = {Portmann, Christopher and Renner, Renato},
  journal      = {Reviews of Modern Physics},
  volume       = {94},
  number       = {2},
  pages        = {025008},
  year         = {2022},
  publisher    = {American Physical Society},
  doi          = {10.1103/RevModPhys.94.025008}
}

@article{bennett1988privacy,
  title        = {Privacy Amplification by Public Discussion},
  author       = {Bennett, Charles H. and Brassard, Gilles and Robert, Jean-Marc},
  journal      = {SIAM Journal on Computing},
  volume       = {17},
  number       = {2},
  pages        = {210--229},
  year         = {1988},
  doi          = {10.1137/0217014}
}

@article{mannalatha2023comprehensive,
  title        = {A comprehensive review of quantum random number generators: concepts, classification and the origin of randomness},
  author       = {Mannalatha, Vaisakh and Mishra, Sandeep and Pathak, Anirban},
  journal      = {Quantum Information Processing},
  volume       = {22},
  pages        = {439},
  year         = {2023},
  publisher    = {Springer},
  doi          = {10.1007/s11128-023-04175-y}
}

@article{li2024improved,
  title         = {Improved Real-Time Post-Processing for Quantum Random Number Generators},
  author        = {Li, Qian and Sun, Xiaoming and Zhang, Xingjian and Zhou, Hongyi},
  journal       = {Advanced Quantum Technologies},
  year          = {2024},
  doi           = {10.1002/qute.202400025},
  eprint        = {2301.08621},
  archivePrefix = {arXiv},
  primaryClass  = {quant-ph}
}

@article{ma2013postprocessing,
  title         = {Postprocessing for Quantum Random-Number Generators: Entropy Evaluation and Randomness Extraction},
  author        = {Ma, Xiongfeng and Xu, Feihu and Xu, He and Tan, Xiaoqing and Qi, Bing and Lo, Hoi-Kwong},
  journal       = {Physical Review A},
  volume        = {87},
  number        = {6},
  pages         = {062327},
  year          = {2013},
  publisher     = {American Physical Society},
  doi           = {10.1103/PhysRevA.87.062327},
  eprint        = {1207.1473},
  archivePrefix = {arXiv},
  primaryClass  = {quant-ph}
}

@article{pironio2010random,
  title        = {Random numbers certified by {B}ell's theorem},
  author       = {Pironio, Stefano and Ac{\'\i}n, Antonio and Massar, Serge and Boyer de la Giroday, A. and Matsukevich, Dmitry N. and Maunz, Peter and Olmschenk, Steven and Hayes, Daniel and Luo, Le and Manning, T. A. and Monroe, Christopher},
  journal      = {Nature},
  volume       = {464},
  number       = {7291},
  pages        = {1021--1024},
  year         = {2010},
  publisher    = {Nature Publishing Group},
  doi          = {10.1038/nature09008},
  url          = {https://www.nature.com/articles/nature09008}
}

@article{carter1979universal,
  title        = {Universal classes of hash functions},
  author       = {Carter, J. Lawrence and Wegman, Mark N.},
  journal      = {Journal of Computer and System Sciences},
  volume       = {18},
  number       = {2},
  pages        = {143--154},
  year         = {1979},
  publisher    = {Elsevier},
  doi          = {10.1016/0022-0000(79)90044-8},
  url          = {https://doi.org/10.1016/0022-0000(79)90044-8}
}

@article{bennett1995generalized,
  title        = {Generalized Privacy Amplification},
  author       = {Bennett, Charles H. and Brassard, Gilles and Cr{\'e}peau, Claude and Maurer, Ueli M.},
  journal      = {IEEE Transactions on Information Theory},
  volume       = {41},
  number       = {6},
  pages        = {1915--1923},
  year         = {1995},
  publisher    = {IEEE},
  doi          = {10.1109/18.476316},
  url          = {https://doi.org/10.1109/18.476316}
}

@inproceedings{renner2005universally,
  title        = {Universally Composable Privacy Amplification Against Quantum Adversaries},
  author       = {Renner, Renato and K{\"o}nig, Robert},
  booktitle    = {Theory of Cryptography Conference (TCC 2005)},
  series       = {Lecture Notes in Computer Science},
  volume       = {3378},
  pages        = {407--425},
  year         = {2005},
  publisher    = {Springer},
  doi          = {10.1007/978-3-540-30576-7_22},
  eprint       = {quant-ph/0403133},
  archivePrefix= {arXiv},
  primaryClass = {quant-ph},
  url          = {https://arxiv.org/abs/quant-ph/0403133}
}

@article{tomamichel2011leftover,
  title        = {Leftover Hashing Against Quantum Side Information},
  author       = {Tomamichel, Marco and Schaffner, Christian and Smith, Adam and Renner, Renato},
  journal      = {IEEE Transactions on Information Theory},
  volume       = {57},
  number       = {8},
  pages        = {5524--5535},
  year         = {2011},
  publisher    = {IEEE},
  doi          = {10.1109/TIT.2011.2158473},
  url          = {https://doi.org/10.1109/TIT.2011.2158473}
}

@article{hayashi2016more,
  title        = {More Efficient Privacy Amplification With Less Random Seeds via Dual Universal Hash Function},
  author       = {Hayashi, Masahito and Tsurumaru, Toyohiro},
  journal      = {IEEE Transactions on Information Theory},
  volume       = {62},
  number       = {4},
  pages        = {2213--2232},
  year         = {2016},
  publisher    = {IEEE},
  doi          = {10.1109/TIT.2016.2526018},
  url          = {https://doi.org/10.1109/TIT.2016.2526018}
}

@article{tsurumaru2022equivalence,
  title        = {Equivalence of Three Classical Algorithms With Quantum Side Information: Privacy Amplification, Error Correction, and Data Compression},
  author       = {Tsurumaru, Toyohiro},
  journal      = {IEEE Transactions on Information Theory},
  volume       = {68},
  number       = {2},
  pages        = {1016--1031},
  year         = {2022},
  publisher    = {IEEE},
  doi          = {10.1109/TIT.2021.3126160},
  eprint       = {2009.08823},
  archivePrefix= {arXiv},
  primaryClass = {quant-ph},
  url          = {https://doi.org/10.1109/TIT.2021.3126160}
}

@article{huang2022stream,
  title={Stream privacy amplification for quantum cryptography},
  author={Huang, Yizhi and Zhang, Xingjian and Ma, Xiongfeng},
  journal={PRX Quantum},
  volume={3},
  number={2},
  pages={020353},
  year={2022},
  publisher={APS}
}

@article{foreman2025cryptomite,
  title={Cryptomite: A versatile and user-friendly library of randomness extractors},
  author={Foreman, Cameron and Yeung, Richie and Edgington, Alec and Curchod, Florian J},
  journal={Quantum},
  volume={9},
  pages={1584},
  year={2025},
  publisher={Verein zur F{\"o}rderung des Open Access Publizierens in den Quantenwissenschaften}
}

@article{ma2016quantum,
  title={Quantum random number generation},
  author={Ma, Xiongfeng and Yuan, Xiao and Cao, Zhu and Qi, Bing and Zhang, Zhen},
  journal={npj Quantum Information},
  volume={2},
  number={1},
  pages={1--9},
  year={2016},
  publisher={Nature Publishing Group}
}

@article{herrero2017quantum,
  title={Quantum random number generators},
  author={Herrero-Collantes, Miguel and Garcia-Escartin, Juan Carlos},
  journal={Reviews of Modern Physics},
  volume={89},
  number={1},
  pages={015004},
  year={2017},
  publisher={APS}
}

@inproceedings{krawczyk1994lfsr,
  title        = {LFSR-based Hashing and Authentication},
  author       = {Krawczyk, Hugo},
  booktitle    = {Advances in Cryptology --- CRYPTO '94},
  series       = {Lecture Notes in Computer Science},
  volume       = {839},
  pages        = {129--139},
  year         = {1994},
  publisher    = {Springer},
  doi          = {10.1007/3-540-48658-5_15}
}

@inproceedings{krawczyk1995new,
  title        = {New Hash Functions for Message Authentication},
  author       = {Krawczyk, Hugo},
  booktitle    = {Advances in Cryptology --- EUROCRYPT '95},
  series       = {Lecture Notes in Computer Science},
  volume       = {921},
  pages        = {301--310},
  year         = {1995},
  publisher    = {Springer},
  doi          = {10.1007/3-540-49264-X_24}
}

@article{araujo2021circulant,
  title        = {The circulant hash revisited},
  author       = {Ara{\'u}jo, Filipe and Neves, Samuel},
  journal      = {Journal of Mathematical Cryptology},
  volume       = {15},
  number       = {1},
  pages        = {250--257},
  year         = {2021},
  publisher    = {De Gruyter},
  doi          = {10.1515/jmc-2018-0054}
}

@inproceedings{carter1977universal,
author = {Carter, J. Lawrence and Wegman, Mark N.},
title = {Universal classes of hash functions (Extended Abstract)},
year = {1977},
isbn = {9781450374095},
publisher = {Association for Computing Machinery},
address = {New York, NY, USA},
url = {https://doi.org/10.1145/800105.803400},
doi = {10.1145/800105.803400},
booktitle = {Proceedings of the Ninth Annual ACM Symposium on Theory of Computing},
pages = {106–112},
numpages = {7},
location = {Boulder, Colorado, USA},
series = {STOC '77}
}

@article{nisan1996randomness,
title = {Randomness is Linear in Space},
journal = {Journal of Computer and System Sciences},
volume = {52},
number = {1},
pages = {43-52},
year = {1996},
issn = {0022-0000},
doi = {https://doi.org/10.1006/jcss.1996.0004},
url = {https://www.sciencedirect.com/science/article/pii/S0022000096900045},
author = {Noam Nisan and David Zuckerman}
}

@article{nie2015generation,
    author = {Nie, You-Qi and Huang, Leilei and Liu, Yang and Payne, Frank and Zhang, Jun and Pan, Jian-Wei},
    title = {The generation of 68 Gbps quantum random number by measuring laser phase fluctuations},
    journal = {Review of Scientific Instruments},
    volume = {86},
    number = {6},
    pages = {063105},
    year = {2015},
    month = {06},
    issn = {0034-6748},
    doi = {10.1063/1.4922417},
    url = {https://doi.org/10.1063/1.4922417}
}

\end{document}